# Magnetic symmetries of terbium tetraboride (TbB$_4$) revealed by resonant x-ray Bragg diffraction

R. D. Johnson[1,2], S. W. Lovesey[2,3,4]

[1] *Department of Physics and Astronomy, University College London, Gower Street, London WC1E 6BT, United Kingdom*

[2]*Diamond Light Source, Harwell Science and Innovation Campus, Didcot, Oxfordshire OX11 0DE, United Kingdom*

[3]*ISIS Facility, STFC, Didcot, Oxfordshire OX11 0QX, United Kingdom*

[4] *Department of Physics, Oxford University, Oxford OX1 3PU, UK*

**Abstract** A recent experimental study of TbB$_4$ at a low temperature using resonant x-ray Bragg diffraction implies a magnetic symmetry not found in any other rare-earth tetraboride. The evidence for this assertion is a change in the intensity of a TbB$_4$ Bragg spot on reversing the handedness (chirality) of the primary x-ray beam [Misawa *et al*., Phys. Rev. B **108**, 134433 (2023)]. It reveals a magnetic chiral signature in TbB$_4$ that is forbidden in phases of rare-earth tetraborides known to date. For, the previous magnetic symmetries are parity-time (PT)-symmetric with anti-inversion present in the magnetic crystal class. Misawa *et al*. appeal to a (PT)-symmetric diffraction pattern to interpret their interesting experimental results. In addition to the use of symmetry that does not permit a chiral signature, calculated patterns impose cylindrical symmetry on Tb sites with no justification. We review magnetic symmetries for TbB$_4$ consistent with a published neutron powder diffraction pattern and susceptibility measurements. On the basis of this information, non-collinear antiferromagnetic order exists below a temperature a temperature $\approx$ 44 K with no ferromagnetic component. Our symmetry-informed patterns encapsulate Tb electronic degrees of freedom in terms of multipoles consistent with established sum-rules for dichroic signals. The investigated symmetry templates are non-centrosymmetric, non-collinear antiferromagnetic constructions with propagation vector **k** = (0, 0, 0). An inferred chiral signature for a parity-even absorption event has an interesting composition. There is the anticipated product of Tb axial dipoles and charge-like quadrupoles (from Templeton-Templeton scattering). Beyond this contribution, though, symmetry allows a product of dipoles in the chiral signature. A predicted change in the intensity of a Bragg spot with rotation of the crystal about the reflection vector (an azimuthal angle scan) can be tested in future experiments. Likewise, contributions to Bragg diffraction patterns from Tb anapoles and higher-order Dirac multipoles.

## I. INTRODUCTION

Rare-earth borides display a raft of interesting electronic properties, including superconductivity and frustrated magnetism [1]. Tetraborides adopt a tetragonal structure and develop collinear or non-collinear antiferromagnetic orders at low temperatures. Terbium tetraboride, of interest here, displays two phase transitions [2]. Neutron powder diffraction patterns for TbB$_4$ are consistent with non-collinear antiferromagnetic order below a temperature $\approx$ 44 K [3]. This type of magnetic order persists beyond a second transition at $\approx$ 24 K, where terbium dipoles tilt toward the a or b axes. Magnetic susceptibility measurements rule against a ferromagnetic component [2], and, reassuringly, there is no trace of it in neutron powder diffraction patterns [3]. Magnetic structures present magnetoelectric and concomitant parity-time (PT)-symmetry [4].

A new magnetic symmetry of TbB$_4$ can be inferred from a Bragg diffraction pattern measured by resonant x-ray diffraction with a sample temperature ≈ 30 K, and it is the principal subject of the present study. A key feature of the pattern reported by Misawa *et al*. is found in their Fig. 4e. It shows that the intensity of the (3, 0, 0) space-group forbidden Bragg spot changes on reversing the chirality of the primary x-rays, from right-handed to left-handed circular polarization, say [5]. Such signatures may be present when magnetic and charge contributions to a diffraction amplitude are separated by a 90º phase shift. Notably, it is forbidden by anti-inversion ($\bar{1}'$) in (PT)-symmetry. For example, magnetoelectric compounds GdB$_4$ (magnetic crystal class 4/m'm'm'), CuMnAs (m'mm) and Cu$_2$(MoO$_4$)(SeO$_3$) (2'/m) do not present chiral signatures [4, 6, 7, 8].

The measured chiral signature Fig. 4e in Misawa *et al*. [5] rules against the magnetic symmetry for a sample temperature ≈ 30 K depicted in their Fig. 1a, since the symmetry elements of magnetic crystal class 4/m'm'm' include anti-inversion ($\bar{1}'$). An appropriate magnetic symmetry for the new phase of TbB$_4$ must also account for the intensity of a space group forbidden Bragg spot displayed in Fig. 3 [5] as a function of the rotation about the reflection vector (an azimuthal angle scan).

We successfully interpret observations on TbB$_4$ reported by Misawa *et al*. [5] with symmetry informed Bragg diffraction patterns. To this end, we use a theory of resonant x-ray Bragg diffraction derived with standard Racah algebra for atomic multipoles [6, 10, 11]. The theory is compatible with tried and tested sum-rules in dichroic signals [12, 13, 14]. This desirable attribute is not fully realized in a phenomenological theory used by Misawa *et al*. that contains free parameters and a constraint to cylindrical Tb site symmetry [15 -17; see footnote #9 in Ref. [15]]. In consequence, diffraction amplitudes are not guaranteed to be compatible with the full magnetic symmetry. Returning to our elected theory, electronic multipoles inferred from experimental data supplied by Misawa *et al*. [5] are listed in Table I. They can be confronted with estimates using an atomic wavefunction for the resonant ion [14, 19, 20], or simulations of electronic structure [21, 22, 23].

## II. TEMPLATE SELECTION

The parent structure of TbB$_4$ is P4/mbm (No. 127, crystal class 4/mmm) with Tb in sites 4g. Suitable templates exclude a ferromagnetic moment [2, 3]. For a second-order phase transition at T$_N$, there is just one template that does not contain anti-inversion, namely, P4'/mbm' (No. 127.391, 4'/mmm'), which transforms as irreducible representation (irrep) $m\Gamma_2^+$. It represents an antiferromagnetic structure with Tb axial dipoles aligned along the crystal c axis. With dipoles aligned in this manor it is no surprise that the diffraction pattern for P4'/mbm' alone does not match data supplied by Misawa *et al*. [5]. We therefore consider linear combinations of irreps, but in this case the magnetic transition must be first-order. Admixing two irreps (more irreps make the search intractable) gives five templates that do not contain anti-inversion. They are P$\bar{4}'$b2' (No. 117.302, $\bar{4}'$m2'), P$\bar{4}'2_1$m' (No. 113.270, $\bar{4}'$2m'), P4'bm' (polar No. 100.174, 4'mm'), P4'$2_1$2' (Sohncke-type No. 90.97, 422' [18]), and Pmc$2_1$ (No. 26.66, 422'), which are associated with the irrep direct sums $m\Gamma_2^+ \oplus m\Gamma_3^-$, $m\Gamma_2^+ \oplus m\Gamma_1^-$, $m\Gamma_2^+ \oplus m\Gamma_4^-$, $m\Gamma_2^+ \oplus m\Gamma_2^-$, and $m\Gamma_2^+ \oplus m\Gamma_5^-$, respectively. We note that all templates include a $m\Gamma_2^+$ mode, and the symmetry adapted modes of magnetic dipoles are depicted for each irrep in Fig. 1a. The reflection condition (h, 0, 0) with h = 2n or equivalent is not met by orthorhombic Pmc$2_1$. The remaining four tetragonal templates are non-centrosymmetric and describe antiferromagnetic non-collinear motifs of Tb dipoles with a propagation vector **k** = (0, 0, 0). The Landau free energy of non-polar templates P$\bar{4}'$b2', P$\bar{4}'2_1$m' and P4'$2_1$2' takes the form [EH + EHH + HEE], where E and H represent electric and magnetic fields, respectively. Magnetoelectric and piezomagnetic effects are allowed. Templates $m\Gamma_2^+ \oplus m\Gamma_1^-$ (P$\bar{4}'2_1$m', No. 113.270, $\bar{4}'$2m') and $m\Gamma_2^+ \oplus m\Gamma_2^-$ (Sohncke-type (chiral) P4'$2_1$2', No. 90.97, 422') are found to match all aspects of the published x-ray diffraction patterns [5]. Corresponding diffraction amplitudes are presented in Section IV. Remaining templates $m\Gamma_2^+ \oplus m\Gamma_3^-$ (P$\bar{4}'$b2', No. 117.302, $\bar{4}'$m2') and $m\Gamma_2^+ \oplus m\Gamma_4^-$ (P4'bm', polar No. 100.174 and an additional field E in the aforementioned Landau free energy, 4'mm') fail with regard to the required chiral signature.

It is likely useful to review the PT-symmetric magnetic structure cited by Misawa *et al*. and rejected by us [5]. The magnetic symmetry depicted in their Fig. 1a for the higher temperature structure that brackets temperatures 24 K and 44 K is a single symmetry adapted mode that transforms by the single irrep $m\Gamma_1^-$. The respective magnetic symmetry is P4/m'b'm' (No.127.395 BNS [9], 4/m'm'm'). The magnetic symmetry depicted in their Fig. 1b for the low temperature symmetry (temperature < 24 K) is a linear combination of of two symmetry adapted modes. One of the two modes is the same as that cited for the above phase observed at a higher temperature. The second mode if appearing by itself presents magnetic symmetry P4'/m'b'm (No. 127.392, 4'/m'm'm). The linear combination drawn in Fig. 1b [5] lowers symmetry to orthorhombic Pb'a'm' (No. 55.359, m'm'm').

### III. RESONANT X-RAY DIFFRACTION

Tuning the energy of x-rays to an atomic resonance has two obvious benefits in diffraction experiments [17, 24]. In the first place, there is a welcome enhancement of Bragg spot intensities and, secondly, spots are element specific. States of x-ray polarization, Bragg angle θ, and the plane of scattering are shown in Fig. 2. A conventional labelling of linear photon polarization states places σ = (0, 0, 1) and π = (cos(θ), sin(θ), 0) perpendicular and parallel to the plane of scattering, respectively [6]. Secondary states σ′ = σ and π′ = (cos(θ), − sin(θ), 0). The x-ray scattering length in the unrotated channel of polarization σ → σ′, say, is modelled by (σ′σ)/D(E). In this instance, the resonant denominator is replaced by a sharp oscillator D(E) = {[E − Δ + iΓ/2]/Δ} with the x-ray energy E in the near vicinity of an atomic resonance $\Delta$ of total width $\Gamma$, namely, $E \approx \Delta$ and $\Gamma \ll \Delta$. The cited energy-integrated scattering amplitude ($\sigma'\sigma$), one of four amplitudes, is studied using standard tools and methods from atomic physics and crystallography. In the first place, a vast spectrum of virtual intermediate states makes the x-ray scattering length extremely complicated [14]. It can be truncated following closely steps in celebrated studies by Judd and Ofelt of optical absorption intensities of rare-earth ions [14, 25-28]. An intermediate level of truncation used here reproduces sum rules for axial dichroic signals created by electric dipole - electric dipole (E1-E1) or electric quadrupole - electric quadrupole (E2-E2) absorption events [6,14]. The attendant calculation presented in Ref. [28] and Section 5.2 in Ref. [6] is lengthy and demanding. Here, we implement universal expressions for scattering amplitudes and abbreviate notation using ($\sigma'\sigma$) $\equiv F_{\sigma'\sigma}$, etc., for amplitudes listed by Scagnoli and Lovesey, Appendix C in Ref. [11]. A similar analysis exists for polar absorption events such as E1-E2 (Appendix D in Ref. [11]), and E1-M1 where M1 is the magnetic moment [29, 30]. Here, we interpret Bragg spots observed by Misawa *et al*. [5] with the x-ray energy tuned to the terbium $L_3$ edge ($E \approx 7.5175$ keV) that accesses E1 ($2p \rightarrow 5d$) and E2 ($2p \rightarrow 4f$) absorption events.

In our adopted description of electronic degrees of freedom, Tb ions are assigned spherical multipoles $\langle O^K_Q \rangle$ of integer rank K with projections Q in the interval $-K \leq Q \leq K$. Angular brackets denote the time-average, or expectation value, of the enclosed spherical tensor operator. A unit-cell electronic structure factor $\Psi^K_Q$ is constructed from all symmetry operations in the chosen space group [9]. Cartesian and spherical components Q = 0, $\pm 1$ of a vector $\mathbf{n} = (\xi, \eta, \zeta)$ are related by $\xi = (n_{-1} - n_{+1})/\sqrt{2}$, $\eta = i(n_{-1} + n_{+1})/\sqrt{2}$, $\zeta = n_0$. A complex conjugate of a multipole is defined as $\langle O^K_Q \rangle^* = (-1)^Q \langle O^K_{-Q} \rangle$, meaning the diagonal multipole $\langle O^K_0 \rangle$ is purely real. The phase convention for real and imaginary parts labelled by single and double primes is $\langle O^K_Q \rangle = [\langle O^K_Q \rangle' + i\langle O^K_Q \rangle'']$. Whereupon, Cartesian dipoles are $\langle O^1_\xi \rangle = -\sqrt{2} \langle O^1_{+1} \rangle'$ and $\langle O^1_\eta \rangle = -\sqrt{2} \langle O^1_{+1} \rangle''$.

Axial (parity even) multipoles denoted $\langle \mathbf{T}^K \rangle$ possess a time signature $(-1)^K$. They can contribute to diffraction enhanced by E1-E1 or E2-E2 absorption events. Bragg spots enhanced by an E1-E1 event are often dominant contributions to a diffraction pattern [16, 17, 24]. All multipoles are functions of the quantum numbers that define the core state of photo-ejected electrons. The dependence on quantum numbers manifests itself in so-called sum rules that relate $\langle \mathbf{O}^K \rangle$ measured at $L_2$ and $L_3$ edges, for example [6, 12, 13, 14]. Sum rules Eqs. (A5) and (A6) for the axial dipole $\langle \mathbf{T}^1 \rangle$ present the orbital angular momentum $\langle \mathbf{L} \rangle$ in the valence state. Dirac atomic multipoles $\langle \mathbf{G}^K \rangle$ are polar (parity odd) and magnetic (time odd) [6, 14]. They are permitted in a magnetic material when the resonant ion occupies an acentric site. Detection of Dirac multipoles requires a probe with matching attributes, of course, which are found in x-ray diffraction enhanced by E1-E2 or E1-M1 parity-odd absorption events.

# IV. CHIRAL SIGNATURES

Our chiral signature $\Upsilon$ is the intensity added to a Bragg spot by circular polarization in the primary beam of x-rays with the result [14, 31],

$$\Upsilon = \{(\sigma'\pi)^*(\sigma'\sigma) + (\pi'\pi)^*(\pi'\sigma)\}''. \tag{1}$$

The four x-ray diffraction amplitudes in Eq. (1) are derived from an electronic structure factor Eq. (A1). In more detail, evaluation of Eq. (A1) requires information about the relevant Wyckoff positions found in the Bilbao table MWYCKPOS for the magnetic symmetry of interest [9]. Site symmetry that might constrain projections Q of a multipole is given in the same table. Wyckoff positions are related by operations listed in the table MGENPOS [9]. Taken together, the two tables provide all information required to evaluate Eq. (A1) and, thereafter, all diffraction amplitudes in Eq. (1) with little more work [11]. Multipoles are presented in a lattice basis that is usually specified relative to the chemical parent lattice of the magnetic symmetry, P4/mbm (No. 127) in our case. The basis provides the relation between Miller indices ($H_o$, $K_o$, $L_o$) for the parent lattice and ($h$, $k$, $l$) for the magnetic symmetry.

For future convenience, we attach the label (i) to the tetragonal template $m\Gamma_2^+ \oplus m\Gamma_1^-$ (P$\bar{4}'2_1$m$'$, No. 113.270). The basis relative to P4/mbm is {(1, 0, 0), (0, 1, 0), (0, 0, 1)}. Terbium ions use Wyckoff positions 4e with symmetry m$'_{-xy}$ that imposes the constraint,

$$\sigma_\pi \sigma_\theta (-1)^K \langle O^K_{-Q} \rangle = \sigma_\pi \sigma_\theta (-1)^{K+Q} \langle O^K_Q \rangle^* = \exp(i\pi Q/2) \langle O^K_Q \rangle, \tag{2}$$

where $\sigma_\pi$ and $\sigma_\theta$ are parity and time signatures, respectively. For parity-even absorption events (E1-E1 and E2-E2) $\sigma_\pi = +1$ and $\sigma_\theta = (-1)^K$, while $\sigma_\pi = -1$ and $\sigma_\theta = -1$ for Dirac multipoles [5, 14]. Axial magnetic dipole magnitudes along the a and b crystal axes in (i) are identical, i.e., $\langle T^1_a \rangle = \langle T^1_b \rangle$, and the magnetic symmetry is depicted in Fig. 1b.

For the space-group forbidden reflection ($h$, 0, 0) with odd $h$ studied by Misawa *et al*. [5], the electronic structure factor Eq. (A2) satisfies $\Psi^K_Q(i) = 0$ using multipoles with even K, Q = 0 and $\sigma_\pi = +1$. In the case of an E1-E1 event,

$$(\sigma'\sigma)_i = 0, \ (\pi'\pi)_i \propto \sin(2\theta) \ [i\alpha' \cos(\psi) \langle T^1_c \rangle + \alpha'' \sin(\psi) \langle T^1_a \rangle], \tag{3}$$

and intensity in the rotated channel of polarization,

$$|(\pi'\sigma)_i|^2 \propto \cos^2(\theta) \ [\{\alpha'' \cos(\psi) \langle T^1_a \rangle - \sqrt{2} \ \alpha' \sin(\psi) \langle T^2_{+2} \rangle''\}^2 \tag{4}$$
$$+ \{\alpha' \sin(\psi) \langle T^1_c \rangle + \sqrt{2} \ \alpha'' \cos(\psi) \langle T^2_{+1} \rangle'\}^2].$$

Subscripts a and c on axial dipoles denote cell edges depicted in Fig. 1. A spatial phase factor $\alpha = \exp(i2\pi hx) = (\alpha' + i\alpha'')$ with x ≈ 0.3172, and $\alpha'/\alpha'' \approx -3.25$ for (3, 0, 0) [3, 5]. Unimportant numerical pre-factors omitted in the foregoing results, and those that follow, explain the use of a proportionality sign. The corresponding chiral signature is,

$$\Upsilon(i) \propto \cos(\theta) \sin(2\theta) [\sqrt{2}\, \alpha' \alpha'' \langle T^1_a \rangle \langle T^1_c \rangle \quad (5)$$
$$+ \sin(2\psi) \{\alpha''^2 \langle T^1_a \rangle \langle T^2_{+1} \rangle' - \alpha'^2 \langle T^1_c \rangle \langle T^2_{+2} \rangle''\}].$$

Crystal axes (a, b, c) and photon axes (x, y, z) in Fig. 2 are correctly aligned for an azimuthal origin $\psi = 0$. The chiral signature $\Upsilon(i)$ is proportional to $[\langle T^1_a \rangle \langle T^1_c \rangle]$ at $\psi = -90°$, and a non-zero signature at this azimuth accords with experimental data [5]. Likewise for intensity in the rotated channel of polarization as a function of $\psi$ reproduced in Fig. 3. The chiral signature $\Upsilon(90.97)$ is the same as Eq. (5) apart from a change in the sign that accompanies $\alpha'^2$. Furthermore, intensity in the rotated channel for No. 90.97 is the same as Eq. (4) after a change in the sign with $\langle T^2_{+2} \rangle''$. Hence, template $P4'2_12'$ (No. 90.97, $422'$) may also account for experimental diffraction intensities [5].

Estimates of the four multipoles in $|(\pi'\sigma)_i|^2$ in the magnetically ordered phase (30 K) of TbB$_4$ are inferred from a fit to experimental data [5]. To this end, we parameterize Eq. (4),

$$[u \cos(\psi)]^2 + [v^2 + w^2] \sin^2(\psi) - uw \sin(2\psi),$$

and deduce $u \approx 0.52$, $v \approx 1.04$ and $w \approx 0.11$ that generate the blue curve in Fig. 3. Table I contains the relative magnitudes of corresponding multipoles as a function of $\langle T^1_a \rangle$ (arbitrary units). Notable features include similar magnitudes for $\langle T^1_a \rangle$ and $\langle T^1_c \rangle$, and an inverse relation between the magnitudes of dipoles and quadrupoles. In so far as Eq. (4) is relevant for the paramagnetic phase (60 K), the red curve in Fig. 3 is generated with $\langle T^2_{+2} \rangle'' \approx 0.63$, and $\langle T^1_a \rangle = \langle T^1_c \rangle = \langle T^2_{+1} \rangle' = 0$. While null values for dipoles are required in the paramagnetic phase, an implied large change in quadrupoles at the magnetic phase transition suggests significant alterations to material properties.

Returning to Sohncke-type $P4'2_12'$ (No. 90.97, enantiomorphous (chiral) crystal class $4'22'$), Tb ions use Wyckoff positions 4e and electronic structure factors $\Psi^K_Q(i)$ in Eq. (A2) and $\Psi^K_Q(90.97)$ are formally identical. However, site symmetries are not the same with $2'_{xy}$ for No. 90.97 and the constraint by $\exp(-i\pi Q/2) \langle O^K_Q \rangle = \sigma_\theta (-1)^K \langle O^K_{-Q} \rangle$. Inversion, mirror, improper rotations and glide symmetries are absent in Sohncke lattices [18]. A neutral screw axis $2_1$ in $P4_2 12$ is achiral while the atomic structure around the axis is chiral. Of the 65 Sohncke lattices primitive ones are chiral and centred ones are not. Orthorhombic and lower symmetry lattices do not contain one of 11 enantiomorphous pairs and the related space groups are achiral.

Moving on, the tetragonal polar template $m\Gamma_2^+ \oplus m\Gamma_4^-$ ($P4'bm'$, No. 100.174), hereafter labelled (ii), possesses a basis $\{(1, 0, 0), (0, 1, 0), (0, 0, 1)\}$ relative to the parent No. 127, i.e., the same basis as for template (i). Terbium ions occupy Wyckoff positions 4c, the constraint Eq. (2) applies, $\langle T^1_a \rangle = \langle T^1_b \rangle$, and the magnetic symmetry is depicted in Fig. 1a. For an E1-E1 event and a reflection vector $(h, 0, 0)$ with odd $h$,

$$(\sigma'\sigma)_{ii} \propto i\alpha'' \sin(2\psi) \langle T^2_{+1} \rangle',$$

$$(\pi'\pi)_{ii} \propto i [\sin(2\theta) \alpha' \cos(\psi) \langle T^1_c \rangle + \sqrt{2} \sin^2(\theta) \alpha'' \sin(2\psi) \langle T^2_{+1} \rangle'],$$

$$|(\pi'\sigma)_{ii}|^2 \propto [\{\sqrt{2} \sin(\theta) \alpha'' \cos(2\psi) \langle T^2_{+1} \rangle' - \cos(\theta) \alpha' \sin(\psi) \langle T^1_c \rangle\}^2$$

$$+ \{\sqrt{2} \cos(\theta) \alpha' \sin(\psi) \langle T^2_{+2} \rangle'' - \alpha'' \sin(\theta) \langle T^1_a \rangle\}^2]. \qquad (6)$$

In contrast to (i), diffraction is allowed in the unrotated channel. Our result for the chiral signature is,

$$\Upsilon(ii) \propto \cos(\theta) \cos(\psi) [\sin(2\theta) \sin(\psi)\{\alpha'^2 \langle T^1_c \rangle \langle T^2_{+2} \rangle'' - \alpha''^2 \langle T^1_a \rangle \langle T^2_{+1} \rangle'\}$$

$$+ \sqrt{2} \alpha' \alpha''\{2 (\cos(\theta) \sin(\psi))^2 \langle T^2_{+1} \rangle' \langle T^2_{+2} \rangle'' - \sin^2(\theta) \langle T^1_a \rangle \langle T^1_c \rangle \}]. \qquad (7)$$

The c axis is normal to the plane of scattering in Fig. 3 at the start of an azimuthal angle scan, the setting used by Misawa *et al*. [5]. Notably, $\Upsilon(ii)$ is proportional to $\cos(\psi)$ and zero for $\psi = -90°$. The result conflicts with supplied measurements, while chiral signatures for templates (i) and No. 90.97 do not.

Use of magnetic symmetry No. 117.302 as a template for magnetic $TbB_4$ is not supported by measurements. Its basis is $\{(0, 1, 0), (-1, 0, 0), (0, 0, 1)\}$ relative to No. 127. Terbium ions are in Wyckoff positions 4g with symmetry $2'_{xy}$. The electronic structure factor is formally identical to Eq. (A2) with an explicit dependence on the parity of electronic multipoles, but the Tb site symmetry is not the same as that for (ii). However, $\Upsilon(117.302)$ is zero for $\psi = -90°$ like $\Upsilon(ii)$.

## V. DIRAC MULTIPOLES

An anapole (Dirac dipole, $\sigma_\pi \sigma_\theta = +1$) depicted in Fig. 4 diffracts x-rays in resonant scattering enhanced by a parity-odd electric dipole-electric quadrupole (E1-E2) event, for example. [A vector product $(\mathbf{R} \times \mathbf{S})$ where $\mathbf{R}$ and $\mathbf{S}$ are electronic space (time-even and polar) and spin (time-odd and axial) variables, respectively, represents the discrete symmetry of a spin anapole.] Experimental results for Dirac multipoles in $V_2O_3$ and CuO have been published together with successful interpretations [14, 32, 33]. In the case of $TbB_4$ available Tb resonance events include $L_3$; $2p \rightarrow 5d$ and $2p \rightarrow 4f$ for E1 and E2, and $M_5$; $2d \rightarrow 4f$ and $2d \rightarrow 5d$ for E1 and E2. Diffraction illuminates Dirac multipoles with ranks K = 1, 2, 3. Energies of E1-E1 and E1-E2 resonances are expected to be different.

We examine the information available for the polar template (ii) using an E1-E2 absorption event. Terbium ions in (ii) support polar multipoles ($\sigma_\pi \sigma_\theta = -1$) at all temperatures, of course. Dirac multipoles ($\langle \mathbf{G}^K \rangle$ with $\sigma_\pi \sigma_\theta = +1$) are also a manifestation of the polar structure and epitomize the low temperature magnetic phase. Anapoles in template (ii) satisfy $\langle G^1_a \rangle = - \langle G^1_b \rangle$ according to Eq. (2). Diffraction amplitudes are derived from Eq. (A3) using expressions in Appendix D of Ref. [11]. Intensity in the rotated channel for ($h$, 0, 0) with odd $h$ is predicted to be,

$$|(\pi'\sigma)|^2 \propto [\{\alpha' \langle G^2_0 \rangle [(\cos(\theta)\cos(\psi))^2 - \sin^2(\theta)]\}^2$$
$$+ (2/15) \{\sin(2\theta)\, \alpha'' \sin(\psi) [3\langle G^1_{+1}\rangle' + 2\sqrt{5}\, \langle G^2_{+1}\rangle'$$
$$+ (4 - 15\cos(\psi)^2) \langle G^3_{+1}\rangle' + \sqrt{15}\, \cos(\psi)^2 \langle G^3_{+3}\rangle']\}^2]. \quad (8)$$

As with an E1-E1 absorption event, the chiral signature is zero for $\psi = -90°$. This finding follows from the E1-E2 result,

$$\Upsilon(\text{ii}) \propto \alpha' \alpha'' \langle G^2_0 \rangle \cos(\theta) \cos(\psi) [(\cos(\theta)\cos(\psi))^2 - \sin^2(\theta)]$$
$$\times [-\sin^2(\theta) \langle G^1_{+1}\rangle' + (2/3)\sqrt{5} \cos(2\theta)\, \langle G^2_{+1}\rangle'$$
$$+ (1/12) \{(4 + \cos^2(\theta))(11 - 15\cos(\psi)^2) - 1\} \langle G^3_{+1}\rangle' \quad (9)$$
$$- (1/4)\sqrt{(5/3)}\{(\cos(\theta)\sin(\psi))^2 + 1\}\langle G^3_{+3}\rangle']. $$

Bragg spots ($h$, 0, 0) with odd $h$ reveal an anapole along the crystal a axis, a quadrupole (K = 2) and an octupole (K = 3).

## VI. CONCLUSIONS AND DISCUSSION

In summary, we confront four templates of magnetic symmetry designed for terbium tetraboride with limited Bragg diffraction data [5]. Designs descend from the parent structure of $TbB_4$ and exclude a ferromagnetic component on the grounds of published susceptibility data [2]. Two of the templates match published resonant x-ray diffraction patterns on two counts [5]. First, a chiral signature that measures the change to the intensity of a space-group forbidden magnetic Bragg spot brought about by reversing the chirality of the primary x-rays, from right-handed to left-handed circular polarization, say. Secondly, a change in intensity with rotation of the crystal about the reflection vector, usually called an azimuthal angle scan. Notably, a non-zero chiral signature relies on the product of two axial dipoles. Conventional thinking about the signature is in terms of products of a magnetic dipole and a charge-like quadrupole from Templeton-Templeton scattering. Indeed, such a result is exploited by Misawa *et al*. without due regard for symmetry constraints; see Eq. (12) in Ref. [5]. To reiterate, the measured chiral signature - the (3, 0, 0) Bragg spot supplied in Fig. 4e in Misawa *et al*. [5]

- rules against a PT-symmetric magnetic structure. The two successful templates have magnetic symmetry $P\bar{4}'2_1m'$ (i, No. 113.270) and $P4'2_12'$ (No. 90.97, 422′), respectively, with the former symmetry encompassing the previously reported in-plane magnetic structure augmented by an antiferromagnetic mode polarised parallel to the c-axis. A polar template $P4'bm'$ (ii, No. 100.174) does not present a chiral signature at the observed azimuthal angle, although it meets success with intensity in the rotated channel of polarization Eq. (6). Diffraction patterns for Dirac multipoles in Section V tell us more about template (ii) with a view to future experiments. Continuing in this vein, the prediction is that the unrotated diffraction amplitude for template (i) is zero whereas it can be different from zero for (ii).

An investigation of a ferromagnetic component in the magnetic symmetry with interesting results has used monoclinic $P2_1/c$ (No. 14.75, magnetic crystal class 2/m). It is centrosymmetric unlike (i) and (ii). The Landau free energy includes [H + HEE], with non-linear magnetoelectric and piezomagnetic effects allowed. The $P2_1/c$ magnetic symmetry is reached through a linear combination of three symmetry-adapted modes defined with respect to the P4/mbm (No. 127) parent. Two of these modes define orthogonal antiferromagnetic (AFM) and ferromagnetic (FM) configurations with magnetic dipoles pointing along a and b (or b and a) axes, respectively. They correspond to magnetic symmetry $Pb'am'$ (orthorhombic No. 55.358, m'mm'). The addition of an AFM mode with moment components along the crystal c axis (magnetic symmetry $P4'/mbm'$, No. 127.391, 4′/mmm′), lowers the symmetry to $P2_1/c$ and non-collinear AFM order. Terbium ions occupy general Wyckoff positions 4e that are devoid of symmetry including spatial inversion. The chiral signature $\Upsilon(P2_1/c)$ in Eq. (B1) for reflections (0, k, 0) with odd k matches $\Upsilon(i)$ in reproducing the limited experimental data.

**APPENDIX A: Electronic Structure Factors**

Our electronic structure factor [6, 14],

$$\Psi^K_Q = [\exp(i\boldsymbol{\kappa}\cdot\mathbf{d})\langle O^K_Q\rangle_\mathbf{d}], \tag{A1}$$

delineates a Bragg diffraction pattern for a reflection vector $\boldsymbol{\kappa}$ defined by integer Miller indices (h, k, l). The implied sum in Eq. (A1) is over Wyckoff positions **d** used by Tb ions. The $\Psi^K_Q$ respects all symmetries of the specified magnetic space group [9]. The basis relative to the parent No. 127 for templates (i) and (ii) are the same with cell dimensions $a \approx 7.120$ Å, $b \approx 7.120$ Å, $c \approx 4.042$ Å, and alpha = beta = gamma = 90°.

For Wyckoff positions 4e in tetragonal $P\bar{4}'2_1m'$ (No. 113.270), the magnetic symmetry labelled (i) in the main text,

$$\Psi^K_Q(i) = (-1)^k \langle O^K_Q \rangle [\alpha\beta + (\alpha\beta)^* (-1)^Q]$$

$$+ (-1)^h \langle O^K_{-Q} \rangle (-1)^K [\alpha\beta^* + \alpha^*\beta (-1)^Q]. \tag{A2}$$

Spatial phase factors are $\alpha = \exp(i2\pi h x)$ and $\beta = \exp(i2\pi k x)$ with a general coordinate $x \approx 0.3172$ [3]. The identity $\Psi^1_Q(i) = 0$ obtained with $\kappa = (0, 0, l)$ is expected, because symmetry in the magnetic crystal class $\bar{4}'2m'$ forbids (bulk) ferromagnetism. Turning to Wyckoff positions 4c in the polar template P4'bm' (No. 100.174), the corresponding electronic structure factor $\Psi^K_Q(ii)$ is similar to Eq. (2). We find,

$$\Psi^K_Q(ii) = (-1)^k \langle O^K_Q \rangle [\alpha\beta + (\alpha\beta)^* (-1)^Q]$$

$$+ \sigma_\pi (-1)^h \langle O^K_{-Q} \rangle (-1)^{K+Q} [\alpha\beta^* + \alpha^*\beta (-1)^Q]. \quad (A3)$$

The general coordinate is again $x \approx 0.3172$, and multipoles in both $\Psi^K_Q(i)$ and $\Psi^K_Q(ii)$ must satisfy Eq. (1). Notably, $\Psi^K_Q(ii)$ depends explicitly on parity via its signature $\sigma_\pi$. The identity for ferromagnetism prevails with $\sigma_\pi = +1$.

Lastly, for Wyckoff positions 4e in monoclinic P2$_1$/c (No. 14.75) mentioned in Section VI and the subject of Appendix B,

$$\Psi^K_Q(P2_1/c) = \langle O^K_Q \rangle [\beta\gamma + \sigma_\pi (\beta\gamma)^*] + (-1)^{k+l}(-1)^{K+Q} \langle O^K_{-Q} \rangle [\beta^*\gamma + \sigma_\pi (\beta^*\gamma)^*]. \quad (A4)$$

The basis relative to the parent No. 127 is {(0, 0, 1), (1, 0, 0), (0, 1, 0)} with cell dimensions $a \approx 4.042$ Å, $b \approx 7.120$ Å and $c \approx 7.120$ Å, and alpha = beta = gamma = 90°. The spatial phase factors are $\beta = \exp(i2\pi k y)$ and $\gamma = \exp(i2\pi l z)$ with general coordinates $y \approx 0.3175$ and $z \approx 0.8175$. The result $\Psi^1_{+1}(P2_1/c) \propto \langle T^1_{+1} \rangle''$ obtained with $\kappa = (h, 0, 0)$ confirms that a ferromagnetic component along the tetragonal crystal a axis is permitted.

The axial dipole $\langle \mathbf{T}^1 \rangle$ in our elected theory of Bragg diffraction obeys a sum rule announced in the first place for dichroic signals [12, 13]. The E1-E1 sum rule presents the orbital angular momentum $\langle \mathbf{L} \rangle_l$ in the valence state. Specifically [6],

$$[\langle \mathbf{T}^1 \rangle_- + \langle \mathbf{T}^1 \rangle_+]_{11} = \langle \mathbf{L} \rangle_l [l_c(l_c + 1) - 2 - l(l+1)] \times [2\sqrt{2}\, l(l+1)(2l+1)]^{-1}. \quad (A5)$$

Here, $l$ and $l_c$ are the valence and core angular momenta, and subscripts $\pm$ denote total core angular momenta ($l_c \pm 1/2$). In consequence, a sum of dipoles at L$_2$ and L$_3$ edges satisfies $[\langle \mathbf{T}^1 \rangle_- + \langle \mathbf{T}^1 \rangle_+]_{11} = -\langle \mathbf{L} \rangle_d/(10\sqrt{2})$ on using states p → d [19]. The dipole sum rule for an E2-E2 event is $[\langle \mathbf{T}^1 \rangle_- + \langle \mathbf{T}^1 \rangle_+]_{22} = -\langle \mathbf{L} \rangle_f (1/21)\sqrt{(2/5)}$ on using p → f in,

$$[\langle \mathbf{T}^1 \rangle_- + \langle \mathbf{T}^1 \rangle_+]_{22} = \langle \mathbf{L} \rangle_l [l_c(l_c + 1) - 6 - l(l+1)] \times [2\sqrt{10}\, l(l+1)(2l+1)]^{-1}. \quad (A6)$$

### APPENDIX B: Monoclinic Template

The descent in lattice symmetry from tetragonal to monoclinic P2$_1$/c is allowed without a change of origin or unit cell volume. The magnetic crystal class 2/m is a subgroup of 4/mmm.

From the electronic structure factor Eq. (A4), an antiferromagnetic motif of Tb axial dipole moments with zero propagation vector $\mathbf{k} = (0, 0, 0)$ can possess components in the tetragonal (bc) plane together with a ferromagnetic component along the a axis. Miller indices for the parent structure $(H_o, K_o, L_o)$ and monoclinic magnetic structure $(h, k, l)$ are related by $h = L_o$, $k = H_o$, $l = K_o$. The $P2_1/c$ chiral signature for reflections $(0, k, 0)$ with odd $k$ is,

$$\Upsilon(P2_1/c) \propto (\beta')^2 \cos(\theta) \sin(2\theta) [\sin(\psi) \langle T^1_b \rangle + \cos(\psi) \langle T^1_c \rangle]$$

$$\times [\cos(\psi) \langle T^2_{+2} \rangle'' - \sin(\psi) \langle T^2_{+1} \rangle'], \tag{B1}$$

with $\beta' = \cos(2\pi k x)$ [3]. The origin of the azimuthal angle $\psi = 0$ posts the tetragonal c axis normal to the plane of scattering in Fig. 3. Evidently, $\Upsilon(P2_1/c)$ can be different from zero for an azimuthal angle $\psi = -90°$, which matches $\Upsilon(i)$ and data in Fig. 4e [5]. The result Eq. (B1) is derived from diffraction amplitudes,

$$(\sigma'\sigma) = 0, \ (\pi'\pi) \propto -i\beta' \sin(2\theta) [\sin(\psi) \langle T^1_b \rangle + \cos(\psi) \langle T^1_c \rangle], \tag{B2}$$

$$(\pi'\sigma) \propto \beta' \cos(\theta) [i\{\sin(\psi) \langle T^1_c \rangle - \cos(\psi) \langle T^1_b \rangle\} + \sqrt{2}\{\cos(\psi) \langle T^2_{+2} \rangle'' - \sin(\psi) \langle T^2_{+1} \rangle''\}].$$

Setting $\langle T^2_{+2} \rangle'' \approx 0$ in the intensity $|(\pi'\sigma)|^2$ brings satisfactory agreement with the experimental results. Template (i) and $P2_1/c$ (No. 14.75) do not allow diffraction in the unrotated channel since $(\sigma'\sigma) = 0$ for both magnetic symmetries.

By way of a contrast in magnetic properties based on monoclinic $P2_1/c$, (PT)-symmetry in $P2_1'/c$ used by $Cu_2(MoO_4)(SeO_3)$ (No. 14.77, 2'/m) forbids a chiral signature [8]. A piezomagnetic effect and ferromagnetism are forbidden, and the Landau free energy [EH] is compatible with a linear magnetoelectric effect. Orthogonal Cartesian axes $(\xi, \eta, \zeta)$ with unique axis $\eta$ may frame an electronic structure factor. Regarding diffraction enhanced by an E1-E1 absorption event, all $(0, 0, 2n + 1)$ amplitudes can be different from zero for example. Active axial multipoles are dipoles $\langle T^1_\xi \rangle$ and $\langle T^1_\zeta \rangle$ in the plane normal to the $\eta$ axis, and quadrupoles $\langle T^2_{+1} \rangle''$ and $\langle T^2_{+2} \rangle''$. The latter appears in the unrotated channel with a diffraction amplitude $(\sigma'\sigma) \propto [\sin(2\psi) \langle T^2_{+2} \rangle'']$, while $[\cos(\psi) \langle T^1_\xi \rangle]$ and $[\sin(2\psi) \langle T^2_{+2} \rangle'']$ feature in $(\pi'\pi)$.

**ACKNOWLEDGEMENTS** SWL is pleased to acknowledge ongoing intellectual support from Dr K. S. Knight.

(a)

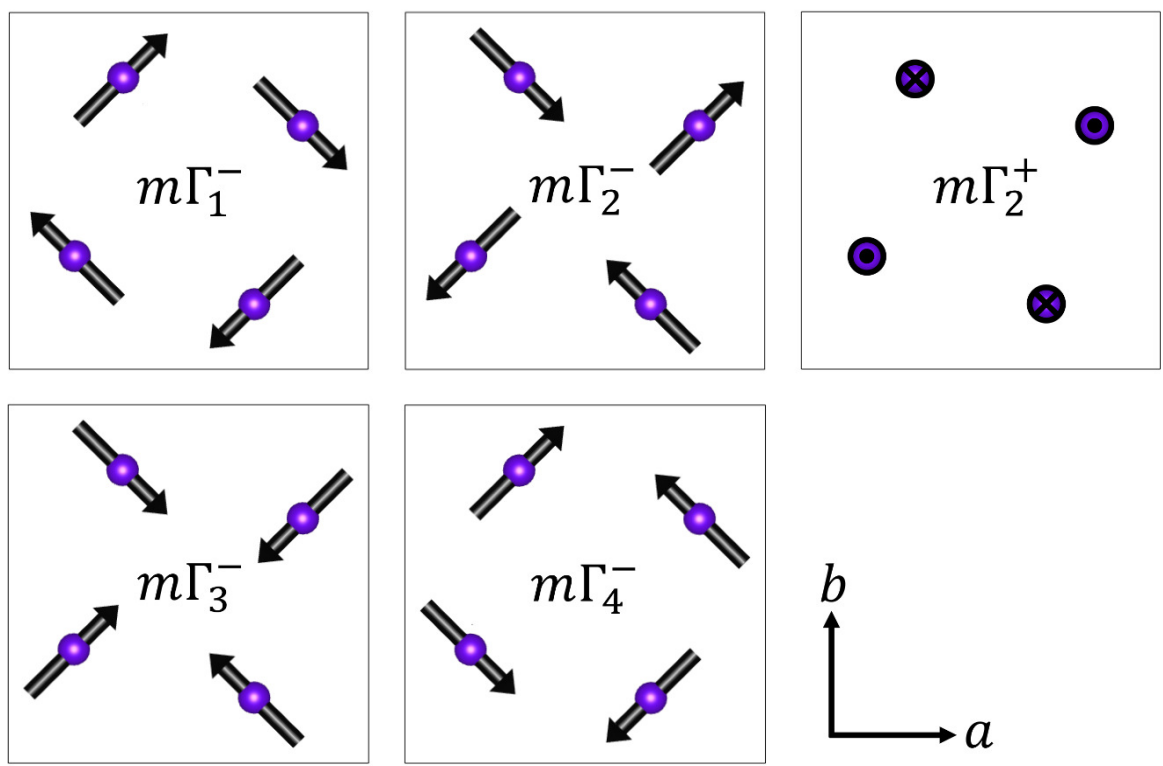

(b)

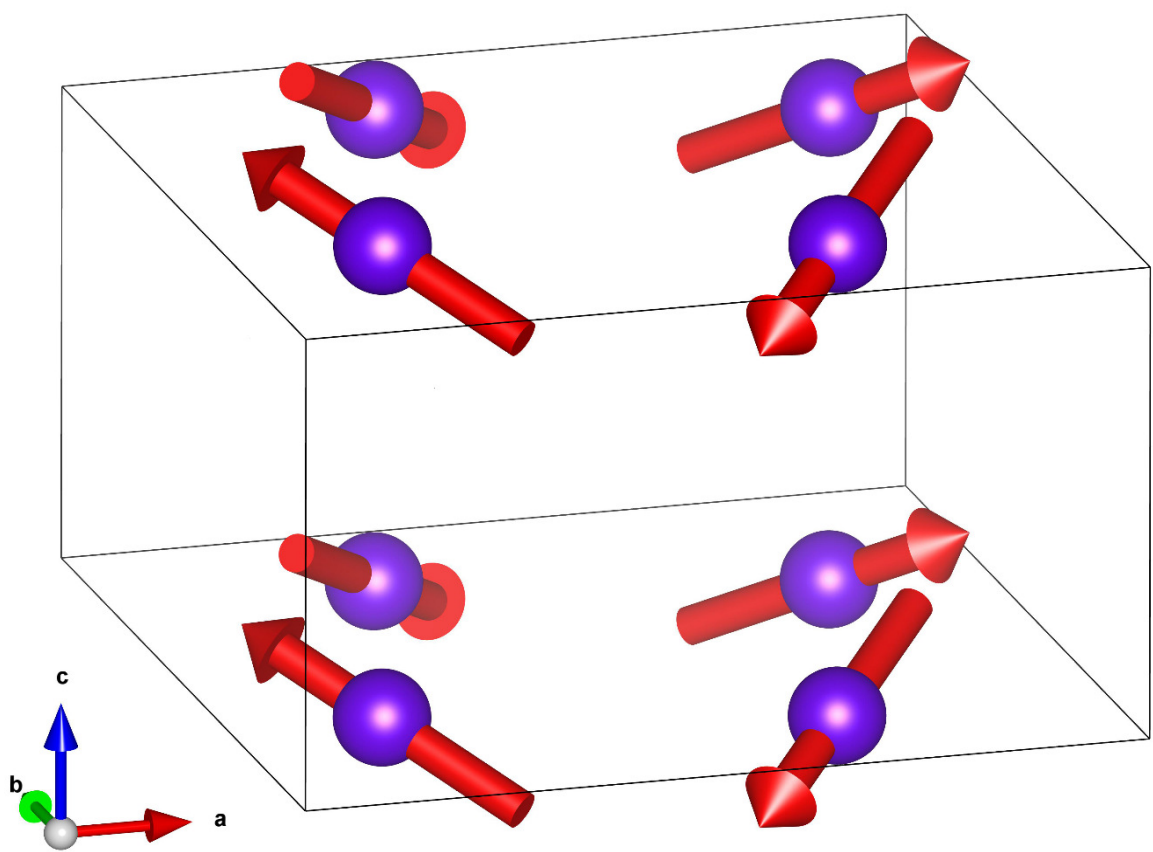

**FIG. 1.** Panel (a); Axial dipole motifs (i) $m\Gamma_2^+ \oplus m\Gamma_1^-$ (P$\bar{4}'2_1$m', No. 113.270), magnetic crystal class $\bar{4}'2m'$) and $m\Gamma_2^+ \oplus m\Gamma_2^-$ (No. 90.97, 422') match diffraction patterns [5]. Panel (b); Depiction of axial dipoles with magnetic symmetry No. 113.270.

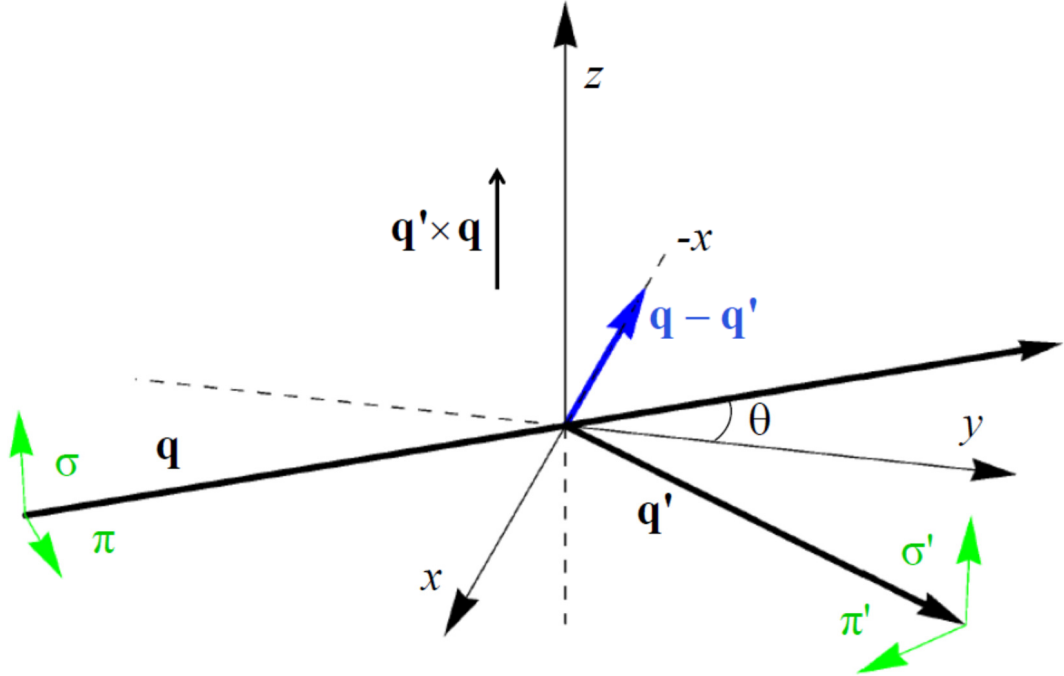

**FIG. 2**. Primary (σ, π) and secondary (σ', π') states of x-ray polarization. Corresponding wavevectors **q** and **q**' subtend an angle 2θ. The Bragg condition for diffraction is met when **κ** = **q** − **q**' coincides with a reflection vector of the reciprocal lattice. Crystal vectors and the depicted Cartesian (x, y, z) coincide in the nominal setting of the crystal. For the origin of an azimuthal angle ψ = 0 we use the choice made by Misawa *et al.* [5].

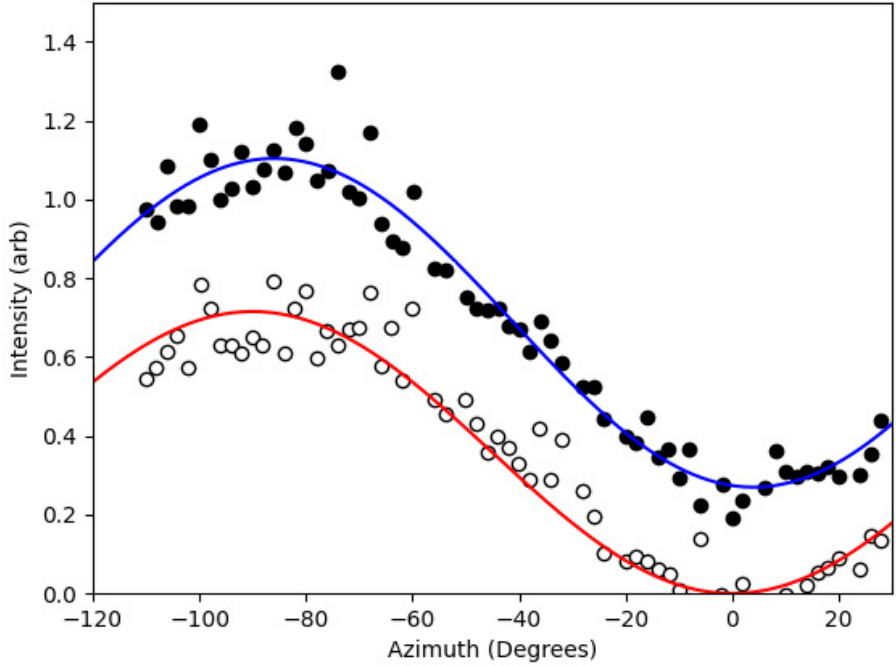

**FIG. 3**. Integrated intensity as a function of azimuthal angle $\psi$ for template (i) with magnetic symmetry No. 113.270. Solid curves are generated using Eq. (4) for intensity in the rotated channel of polarization $|(\pi'\sigma)_i|^2$ and the space-group forbidden reflection (3, 0, 0). Sample temperatures 30 K (magnetically ordered phase) and 60 K (paramagnetic) are labelled blue and red, respectively. Experimental data for TbB$_4$ extracted from Misawa *et al.* (solid points 30 K, open points 60 K) [5].

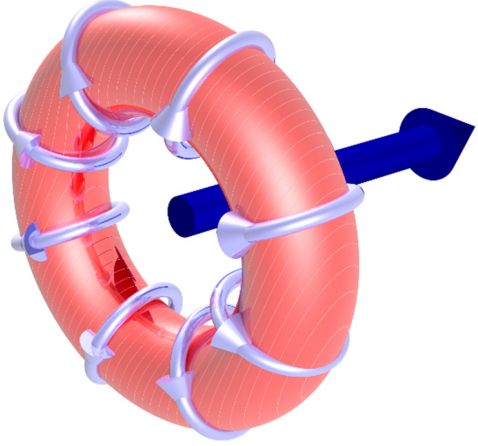

**FIG. 4.** Depiction of a Dirac dipole (an anapole) created by V. Scagnoli.

**TABLE I**. Multipoles for the magnetically ordered phase of TbB$_4$ (30 K) inferred from experimental data depicted in Fig. 3 [5]. Eq. (4) is intensity in the rotated channel of polarization $|(\pi'\sigma)|^2$ for template (i) (magnetic symmetry No. 113.270). A fit to the displayed data as a function of the azimuthal angle $\psi$ is depicted in blue. Corresponding multipoles are listed here as a function of the axial dipole $\langle T^1_a \rangle$ (arbitrary units). It has a maximum value $\langle T^1_a \rangle$ = 1.77 when $\langle T^2_{+2} \rangle'' = 0$.

| $\langle T^1_a \rangle$ | $\langle T^1_c \rangle$ | $\langle T^2_{+1} \rangle'$ | $\langle T^2_{+2} \rangle''$ |
|---|---|---|---|
| 1.7 | 1.08 | 0.34 | 0.13 |
| 1.5 | 0.99 | 0.66 | 0.34 |
| 1.3 | 0.88 | 0.85 | 0.46 |
| 1.1 | 0.77 | 0.98 | 0.55 |
| 0.9 | 0.65 | 1.08 | 0.62 |
| 0.3 | 0.30 | 1.23 | 0.75 |